\documentclass[prl,twocolumn,amssymb,showpacs]{revtex4}

\usepackage{graphicx}
\usepackage{color}
\usepackage{epsfig}
\usepackage{latexsym}
\usepackage{bm}

\begin{document}

\renewcommand{\ni}{{\noindent}}
\newcommand{\dprime}{{\prime\prime}}
\newcommand{\be}{\begin{equation}}
\newcommand{\ee}{\end{equation}}
\newcommand{\bea}{\begin{eqnarray}}
\newcommand{\eea}{\end{eqnarray}}
\newcommand{\nn}{\nonumber}
\newcommand{\bk}{{\bf k}}
\newcommand{\bQ}{{\bf Q}}
\newcommand{\q}{{\bf q}}
\newcommand{\s}{{\bf s}}
\newcommand{\bN}{{\bf \nabla}}
\newcommand{\bA}{{\bf A}}
\newcommand{\bE}{{\bf E}}
\newcommand{\bj}{{\bf j}}
\newcommand{\bJ}{{\bf J}}
\newcommand{\bs}{{\bf v}_s}
\newcommand{\bn}{{\bf v}_n}
\newcommand{\bv}{{\bf v}}
\newcommand{\la}{\langle}
\newcommand{\ra}{\rangle}
\newcommand{\dg}{\dagger}
\newcommand{\br}{{\bf{r}}}
\newcommand{\brp}{{\bf{r}^\prime}}
\newcommand{\bq}{{\bf{q}}}
\newcommand{\hx}{\hat{\bf x}}
\newcommand{\hy}{\hat{\bf y}}
\newcommand{\bS}{{\bf S}}
\newcommand{\cU}{{\cal U}}
\newcommand{\cD}{{\cal D}}
\newcommand{\bR}{{\bf R}}
\newcommand{\pll}{\parallel}
\newcommand{\sumr}{\sum_{\vr}}
\newcommand{\cP}{{\cal P}}
\newcommand{\cQ}{{\cal Q}}
\newcommand{\cS}{{\cal S}}
\newcommand{\ua}{\uparrow}
\newcommand{\da}{\downarrow}
\newcommand{\red}{\textcolor {red}}

\def\lsim {\protect \raisebox{-0.75ex}[-1.5ex]{$\;\stackrel{<}{\sim}\;$}}
\def\gsim {\protect \raisebox{-0.75ex}[-1.5ex]{$\;\stackrel{>}{\sim}\;$}}
\def\lsimeq {\protect \raisebox{-0.75ex}[-1.5ex]{$\;\stackrel{<}{\simeq}\;$}}
\def\gsimeq {\protect \raisebox{-0.75ex}[-1.5ex]{$\;\stackrel{>}{\simeq}\;$}}

\today

\title{Interacting particles in a periodically moving potential: Traveling 
wave and transport}

\author{ Rakesh Chatterjee{$^1$}, Sakuntala Chatterjee{$^{2,\ast}$}, Punyabrata 
Pradhan{$^{2,\ast}$}, and S. S. Manna{$^2$} }

\affiliation{ $^1$CMP Division, Saha Institute of Nuclear Physics, 1/AF Bidhan 
Nagar, Kolkata 700064, India \\ $^2$Department of Theoretical Sciences, S. N. 
Bose National Centre for Basic Sciences, Block - JD, Sector - III, Salt Lake, 
Kolkata 700098, India }

\begin{abstract}
\noindent{We study a system of interacting particles in a periodically 
moving external potential, within the simplest possible description of 
paradigmatic symmetric exclusion process on a ring. The model describes 
diffusion of hardcore particles where the diffusion dynamics
is locally modified at a uniformly moving defect site, mimicking the effect
 of the periodically moving external potential. The model, though simple, 
exhibits remarkably rich features in particle transport, such as
polarity reversal and double peaks in particle current upon variation of
 defect velocity and particle density. By tuning these variables, the 
most efficient transport can be achieved in either direction along the 
ring. These features can be understood in terms of a traveling 
density wave propagating in the system.
Our results could be experimentally tested, e.g., in a system of colloidal
particles driven by a moving optical tweezer.
}

\end{abstract}

\pacs{05.70.Ln, 05.40.-a, 05.60.-k, 83.50.Ha}

\maketitle

%\section{Introduction}

The advent of state-of-the-art technique to maneuver 
colloidal particles using laser field has opened up new 
avenues of research \cite{Ashkin, Libchaber, Bechinger_PRL2010,
RecentWorks, Reviews}. Recently, single colloidal particle in a 
periodically moving optical potential has been used to experimentally 
investigate several
important aspects \cite{Bechinger_PRL2006, Bechinger_PRL2007, 
Ciliberto_PRL2007, Ciliberto_PRL2009} of nonequilibrium systems.
However, far less explored is the situation where colloidal particles, 
subjected to such a time-periodic potential, can also interact with each other. 
Though the crucial role of interaction
has been studied intensively in the past for nonequilibrium steady states
\cite{Review_Zia, Bechinger_PRL2013, Bechinger2_PRL2013}, not much is 
known about driven many-particle systems having a time-periodic steady state 
where macroscopic properties are a periodic function of time.

In this paper, we ask what happens when a system of interacting colloidal 
particles is driven by 
time-periodic forces. Do the particles always show directional 
motion? What are the conditions for the optimum transport? These questions 
are relevant not only to the colloidal particles, but are also important in 
the context of stochastic pumps \cite{dhar07, Jarzynski_PRL2008, 
Chernyak_PRL2008} and thermal ratchets \cite{MarchesoniPRL1996, 
Seifert_PRL2011, Reimann_2002, Imparato_PRL2012} as well as 
driven fluids in general 
\cite{Penna_Tarazona_JChemPhys2003, Tarazona_Marconi_JchemPhys2008}, e.g., 
in micro-fluidic devices \cite{Quake_RMP2005} manipulated 
by time-varying forces.

\begin{figure}[h]
\begin{center}
\includegraphics[width=6cm]{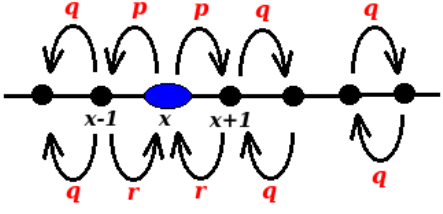}
\caption{(Color online) Schematic diagram of the model. At a particular time,
 the defect site is at
$x$ and marked by an oval shape. Other sites are shown as black solid circles.
Transition from (to) the defect site takes place with rate $p$ (rate $r$). 
All other transitions take place with rate $q$. 
The transition from a particular 
site takes place only if the site is occupied and the destination site is 
empty. }
\label{model_fig}
\end{center}
\end{figure}

We address these questions in a setting of paradigmatic exclusion 
processes \cite {Liggett} where we consider the simplest possible
interaction among the particles, viz. hardcore repulsion, which is present 
in almost all systems due to excluded volume effects
and especially important for dense packing of particles.
The model is defined on a one dimensional periodic lattice of $L$ sites,
each of which can be occupied by at most one particle. 
The effect of a periodically moving external potential 
is modeled as a moving disorder or defect with the diffusive dynamics
modified locally at the defect site which travels along the 
lattice with a uniform velocity $v$ and with a residence time $\tau = 1/v$ at 
each site. A particle hops to its empty nearest neighbor with the following 
rates: (i) $p/2$ from the defect site, (ii) $r/2$ to the defect site and
   (iii) $q/2$ otherwise (see Fig. \ref{model_fig}).
A configuration of the system is specified by occupancy $\{\eta_i 
^{(\alpha)}\}$ of each site and position of the defect site $\alpha$ at
a given time, where the occupancy $\eta_i^{(\alpha)}$
of the $i$-th site takes the value 1 (0) if the site is occupied (empty).

Despite its apparent simplicity, the model exhibits strikingly rich transport
properties arising solely due to the hardcore exclusion among the particles. 
Since time-averaged applied force, due to the external potential, at any 
site is zero, it is not a priori 
clear if the system can support a current and, if so, in which direction.
We find that there is indeed a nonzero current and remarkably the current 
reverses its direction and even shows positive and negative peaks, as 
the defect velocity $v$ and the particle density $\rho$ are varied. 
By tuning $v$ and $\rho$, the most efficient transport can 
be achieved in either direction along the ring. 
Interestingly, the moving defect gives rise to a traveling
wave density pattern in the system, which however always travels in the 
direction of the defect movement. Unlike the perturbative approach used in
\cite{dhar07,dhar08,dhar11}, we consider the case when the disorder
is strong. In this limit, our analytical theory predicts the exact structure 
of the density wave, which explains the above results.

For $v=0$, the model describes an equilibrium system with an external 
potential $V_0$ present only at the defect site. From the detailed balance 
condition, the density at the defect site $\sim \exp(-\beta V_0)$ where 
$V_0 = \beta^{-1}\ln(p/r)$ and $\beta$ inverse temperature. The densities 
at the other sites are uniform. An infinitely large potential barrier 
corresponds to $r=0$.

For nonzero defect velocity, we consider a strongly driven system with 
a large potential barrier where the relaxation dynamics down (up) the 
barrier is much faster (slower) than that in the bulk and also the 
defect velocity is much larger than the bulk relaxation rate, implying 
$p, v \gg q \gg r$. For simplicity, we throughout consider infinite 
barrier, i.e., $r=0$ and, without any loss of generality, one can set 
$p=1$. 

%\section{Mean Field Theory for ${\bm q=0}$}

We first consider the limit $q=0$ which 
means that at any given time, the particle can move if and only if its 
position coincides with the position of the defect site at that time 
\footnote{For small $v$ and $q=0$, an interesting
connection exists between the model and a symmetric exclusion process
with {\it site-wise} ordered sequential update \cite{prep} where
sites are updated consecutively one after another in a particular direction 
along the ring. Note that this ordered sequential update rule 
is different from those studied in the past \cite{Rajewsky_etal_JSP1998}.}.
 This limit is important since, 
as we see later, it provides insights into the case with nonzero $q$.
Starting from a random initial configuration, the system eventually settles 
into a time-periodic steady state where the density-profile has the form 
of a traveling wave moving around the system with the same speed $v$ 
as that of the defect. From now on, unless stated otherwise, we carry 
out all our measurements precisely at the time-steps $t=n \tau$ when the 
moving defect, after spending the residence time $\tau$ at a 
site, is about to move on to the next one, with $n=0,1, \dots \infty$. 
Note that it is not easy to determine the time-periodic steady state for all 
time $t$. However, the analysis becomes much simpler for time $t=n \tau$
when one writes down the following descrete-time evolution equation for density 
$\rho^{(\alpha)}_i(t) = \langle \eta^{(\alpha)}_i(t) \rangle$,
\be
\langle \rho^{(\alpha+1)}(t+\tau)| = \langle \rho^{(\alpha)}(t)| 
{\cal W}^{(\alpha+1)}. \label{evolution1} 
\ee
Here $\langle \rho^{(\alpha)}(t)| \equiv \{\rho^{(\alpha)}_1(t), \dots 
\rho_i^{(\alpha)}(t), \dots, \rho^{(\alpha)}_L(t) \}$ is a row-vector of 
length $L$, with $i$-th element being $\rho_i^{(\alpha)}(t)$ and 
${\cal W}^{(\alpha)}$ is the transition matrix with the defect site at 
$\alpha$. 
   The conditional probability that, given the defect site is occupied, the
   site exchanges particle with its right (left) neighbor during the
   time-interval $\tau$ is denoted as $a_+$ ($a_-$).
   For $q=r=0$, non-vanishing transition rates are found only at the site
   $\alpha$ and hence one can explicitly construct the transition matrix 
in terms of $a_+$ and $a_-$. We have an
   expression for these quantities starting from the microscopic dynamics:
\small
\bea
a_{+} = \left[ \frac{\langle \eta^{(\alpha)}_{\alpha} 
\eta^{(\alpha)}_{\alpha+1} (1-\eta^{(\alpha)}_{\alpha+2})\rangle}{\kappa_1 
\langle \eta^{(\alpha)}_{\alpha+1}\rangle} + \frac{\langle 
(1-\eta^{(\alpha)}_{\alpha})\eta^{(\alpha)}_{\alpha+1}
(1-\eta^{(\alpha)}_{\alpha+2}) \rangle}{2 \kappa_2 \langle 
\eta^{(\alpha)}_{\alpha+1}\rangle} \label{a+} \right] \\
a_{-} = \left[ \frac{\langle (1-\eta^{(\alpha)}_{\alpha}) 
\eta^{(\alpha)}_{\alpha+1} \eta^{(\alpha)}_{\alpha+2} \rangle}{\kappa_1 
\langle \eta^{(\alpha)}_{\alpha+1}\rangle} + \frac{\langle 
(1-\eta^{(\alpha)}_{\alpha}) \eta^{(\alpha)}_{\alpha+1}
(1-\eta^{(\alpha)}_{\alpha+2}) \rangle }{2 \kappa_2 
\langle \eta^{(\alpha)}_{\alpha+1}\rangle}  \right]
\label{a-}
\eea
\normalsize
where the density at the defect site $\alpha$ decays, following a Poisson 
process with rates $1/\kappa_1(v)=[1-\exp(-p/2v)]$ and $1/\kappa_2(v)=
[1-\exp(-p/v)]$, depending on the occupancy of the neighboring sites 
(see Appendix A for details). 
The transition matrix can now be written as 
\bea
{\cal W}^{(\alpha+1)}_{i,j}&=&(1-a_{+}-a_{-}) \mbox{~~~~~~for $i=j=\alpha+1$,}
\nonumber \\
{\cal W}^
{(\alpha+1)}_{i,j}&=&1 \mbox{~~~~~~for $i=j \ne \alpha+1$,}
\nonumber \\
{\cal W}^{(\alpha+1)}_{i,j}&=&a_{+} \mbox{~~~~~~for $i=\alpha+1$ and 
$j=\alpha+2$,}
\nonumber \\
{\cal W}^{(\alpha+1)}_{i,j}&=&a_{-} \mbox{~~~~~~for $i=\alpha+1$ and 
$j=\alpha$.} \nonumber
\eea

In the long time limit, the time-periodic structure of the steady state 
implies that the density profile comes back to itself after each complete 
cycle of the defect movement around the ring, i.e.,  
${\cal W}^{(\alpha)} {\cal W}^{(\alpha+1)} \dots 
{\cal W}^{(L)} {\cal W}^{(1)} \dots {\cal W}^{(\alpha-1)}$ has an eigenvector
$\langle \rho^{(\alpha)}_{st}|$, with eigenvalue unity. Then the $i$th element 
$\rho_{st,i}^{(\alpha)}$ of $\langle \rho^{(\alpha)}_{st}|$, i.e., 
steady-state density at site $i$, satisfies
\be 
\rho_{st,i}^{(\alpha+1)}=\rho_{st,i-1}^{(\alpha)}.
\label{condition1}
\ee
To solve for the density profile in the time-periodic steady state, we note 
that, at the time of
measurement, the defect site $\alpha$ registers a lower density compared to 
the bulk because, for $r=0$, particles cannot hop in to the defect 
site but can only hop out. Similarly, as $q=0$, the neighboring sites 
$(\alpha \pm 1)$ can only receive particles from the defect site but 
they cannot lose particles. The site $(\alpha +1)$ thus has a density 
higher than that at the bulk.
On the other hand, the site $(\alpha-1)$, which could have only lost a particle 
in the previous time step when the defect was at $\alpha-1$, now can receive a 
particle from the defect site $\alpha$ and brings its density back to the bulk level. Therefore, regarding the structure of the density profile as a function of position, 
we formulate an ansatz in the form of a traveling density wave which moves with the 
defect site $\alpha$: 
\bea
\rho_{st, i}^{(\alpha)}&=&\rho_- \mbox{~~~~~~for $i=\alpha$,} 
\nonumber \\
\rho_{st,i}^{(\alpha)}&=&\rho_+ \mbox{~~~~~~for $i=\alpha+1$}
\nonumber \\
\rho_{st,i}^{(\alpha)}&=&\rho_b \mbox{~~~~~~otherwise.} 
\eea
For example, $\langle \rho_{st}^{(1)}| = \{\rho_-, \rho_{+}, \rho_{b}, \dots, 
\rho_{b} \}$ for $\alpha =1$. The ansatz can be used in Eqs. \ref{evolution1} 
and \ref{condition1}, to obtain  
\bea \rho_{_+} a_{+}  + \rho_{_b} = \rho_{_+},
\label{condition2} 
\\ 
\rho_{_+} a_{-} + \rho_{-} = \rho_{_b},
\label{condition2a}
\eea
which can be solved by using particle-number conservation 
$\rho_{_+} \rho_{_-} + (L-2)\rho_{_b} = L \rho$ to get the exact densities 
 \bea 
\rho_{_b} = \frac{(1-a_+)L}{2-a_+-a_-+(1-a_+)(L-2)} \rho \simeq \rho, \\
\rho_{_+} = \frac{1}{1-a_+} \rho_{_b} \simeq \frac{1}{1-a_+} \rho, 
\label{den1} \\ 
\rho_{_-}= \frac{1-a_+-a_-}{1-a_+} \rho_{_b} \simeq 
\frac{1-a_+-a_-}{1-a_+} \rho,
\label{den2}
\eea 
as $L\gg1$.
Note that, we have obtained the density profile in terms of $a_\pm(\rho,v)$ 
which depend on the global density $\rho$ and the defect velocity $v$, and 
involve three-point correlations as in Eqs. \ref{a+} and \ref{a-}. 
From Eqs. \ref{den1} and \ref{den2}, it immediately follows that 
$\rho_{_+}>\rho$ and $\rho_{_-}<\rho$, i.e., a bump and a trough are formed 
respectively in front of the defect site and at the defect site. 
In a many-particle system, due to the lack of closure in time-evolution 
equations for correlation functions (BBGKY hierarchy), it is often difficult 
to obtain such a general structure of the density as a function of position.
Therefore, it is quite remarkable that we obtain an exact 
structure of the density profile for this system. Interestingly, using a 
dynamic density functional theory, a traveling density wave of similar 
structure has been found in a system driven by a moving external 
potential \cite {Penna_Tarazona_JChemPhys2003, Tarazona_Marconi_JchemPhys2008}.

To obtain the current, we note that only the two bonds adjacent
to the defect site can contribute to the current, since 
no hopping takes place across any other bond. As the defect visits 
a particular site with rate $v/L$ the current is
$J_0(\rho, v) = \frac{v}{L} \langle \eta_{\alpha +1}^{(\alpha)} \rangle
(a_+-a_-)$ which can be written in terms of $\rho_\pm$,
\be
J_0(\rho, v) = \frac{v}{L} (\rho_+ + \rho_- - 2\rho),
\label{Jq0}
\ee
after inverting Eqs. \ref{den1}, \ref{den2} and substituting $\langle 
\eta_{\alpha +1}^{(\alpha)} \rangle=\rho_+$.
   The current is nonzero in general as $a_+ \neq a_-$ or
$(\rho_+ - \rho) \neq (\rho - \rho_-)$ from Eqs.
\ref{condition2} and \ref{condition2a}.

So far, we have only discussed the general properties of density
profile and current using exact expressions.
Now we obtain explicit functional dependence of $a_{\pm}(\rho,v)$ on 
$\rho$ and $v$ within mean-field 
theory, where the three point correlations in  Eqs. \ref{a+} and \ref{a-}  
are assumed to be factorized. Therefore, we get
\bea
&a_+ = (1-\rho) \left[\rho_- (1-e^{-p/2v}) + \frac{(1-\rho_-)(1-e^{-p/v})}{2}
\right] \label{MFa+}
\\
&a_- = (1-\rho_-) \left[\rho (1-e^{-p/2v}) + \frac{(1-\rho)(1-e^{-p/v})}{2}
\right] \label{MFa-}
\eea 
Using the above mean field expression for $a_\pm$ into  Eqs. \ref{den1} and 
\ref{den2}, the following quadratic equation can be obtained for $\rho_-$,
\bea
\nonumber
(\rho_- -\rho)\{1-
(1-\rho)(\rho_- \omega_1+(1-\rho_-)\omega_2) \} \\
+ \rho(1-\rho_-)\{\rho \omega_1 + (1-\rho) \omega_2\}=0
\eea 
where $\omega_1 = 1/\kappa_1(v) =[1-\exp(-p/2v)]$ and $ \omega_2 = 1/2\kappa_2(v)=[1-\exp(-p/v)]/2$. 
Out of the two possible solutions, only one is physically relevant as the
other one is larger than unity. The solution for $\rho_-$ now can be 
used, in Eqs. \ref{MFa+} and \ref{MFa-}, to find $a_\pm$ and then $\rho_+$
from Eq. \ref{den1}.
The solutions for $\rho_\pm$ take a particularly simple form in the limit of 
large $v$ where we expand $\omega_1$ and $\omega_2$ in the leading order of 
$1/v$ to obtain 
\bea
\rho_+ = \frac{2 v \rho}{2v-(1-\rho)p} \label{r+_lv} \\
\rho_-= \frac{\rho (2v-2p+p\rho)}{2v-p } \label{r-_lv}
\eea

The mean-field expression of current can be obtained, using Eqs. \ref{MFa+} 
and \ref{MFa-}, as
\be
J_0(\rho, v) = \frac{v}{L}
(1-e^{-p/2v})\rho_+ (\rho_- - \rho).
\label{JMFq0}
\ee  
Interestingly, the current always flows in the direction opposite to the 
defect movement since $\rho_{-}<\rho$. This counter-intuitive result 
can be qualitatively explained in
the following way. The positive current $\rho_+(1-\rho)$ across the bond 
$(\alpha,\alpha+1)$
and the negative current $\rho_+(1-\rho_-)$ across the bond $(\alpha-1,\alpha)$
is due to the diffusive flux from the bump to the bulk and from the bump 
to the trough, respectively. Clearly, the net current is negative.
As shown later, this feature survives even for generic $q$ and $\rho$.
Substituting the previously obtained expressions for $\rho_+$ and $\rho_-$
in Eq. \ref{JMFq0}, the current can be written as a function of $\rho$ and $v$.

\begin{figure}
\begin{center}
\leavevmode
\includegraphics[width=8.5cm,angle=0]{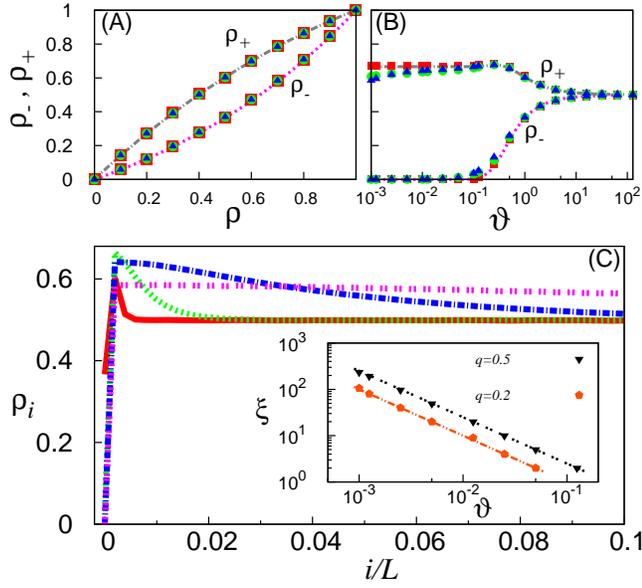}
\caption{(Color online) $\rho_+$ and $\rho_-$ are plotted against global
 density 
$\rho$ with $v=1$ in panel (A) and against the defect 
velocity $v$ with $\rho=0.5$ in panel (B), for $q=0$ (red squares), $0.2$
(green circles) and $0.5$ (blue triangles). Analytical 
prediction for $\rho_+$ ($\rho_-$) is shown by grey (pink) dashed-dotted
(dotted) lines. In panel (C), densities from simulations are plotted against 
scaled distance (scaled by a factor $1/L$) from the defect site for $v=1.0$
 (red solid), $0.1$ 
(green single-dotted), $0.01$ (blue dashed-dotted) and $0.001$ (pink 
double-dotted)
with $q=0.5$. Inset in panel (C) shows the variation of 
$\xi$ against $v$ for $q=0.2$ and $0.5$ (dotted lines - analytical 
predictions). Throughout we use $L=512$, $p=1$, $r=0$.  
}
\label{rho_pm}
\end{center}
\end{figure}

\begin{figure}
\begin{center}
\leavevmode
\includegraphics[width=8.7cm,angle=0]{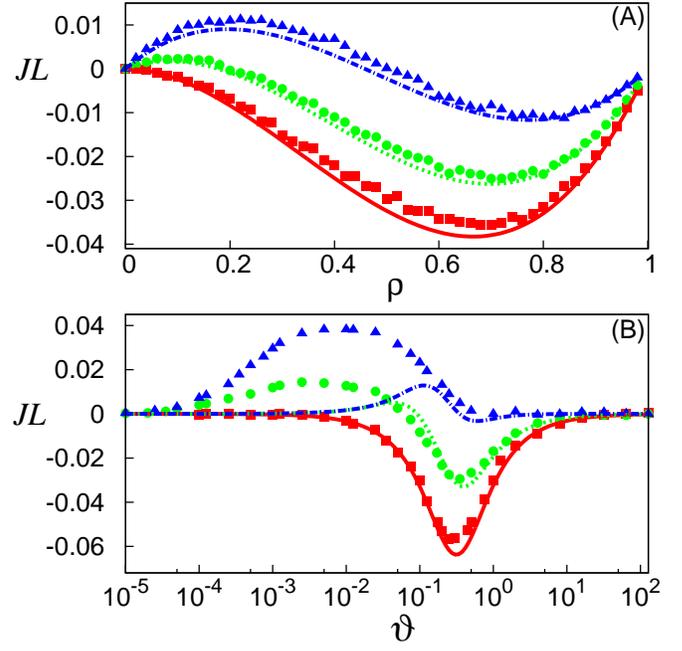}
\caption{(Color online) Scaled current (scaled by a factor $L$)
 is plotted against density $\rho$ for $v=1$ 
in panel (A) and against defect velocity $v$ for $\rho=0.5$ in panel (B) 
for $q=0$ (red squares), $0.2$ (green circles), 
$0.5$ (blue triangles); in both the panels, $p=1$, 
$r=0$ and $L=512$. Analytical results for $q=0,0.2$ and $0.5$ are shown by red
solid, green dotted and blue dashed-dotted lines, respectively.}
\label{j_rho}
\end{center}
\end{figure}

To check the above analytical results, we perform Monte Carlo simulations with $p=1$ (see Appendix B for details). We show the variation 
of $\rho_{\pm}$ as a function of $\rho$ and $v$ in Figs. \ref{rho_pm}(A) 
and \ref{rho_pm}(B), respectively, for $q=0$ (red squares). The
analytical results (lines) show excellent agreement with the simulations. 
We present simulation results (red squares) for current as a 
function of $\rho$ and $v$ in Figs. \ref{j_rho}(A) and \ref{j_rho}(B), 
respectively, again in good agreement with analytical results (red solid line).
Expectedly, for very low and high densities, the current is vanishingly small 
for any finite $v$. The current reaches a negative peak at an intermediate 
density, different from half filling, thus manifesting the absence of 
particle-hole symmetry. Similar non-monotonic variation of current is 
observed as $v$ is varied for a fixed $\rho$.  
For small $v \ll 1$, $\rho_{\pm}$ are independent of 
$v$ and therefore current $J_0 \sim v$. For large $v \gg 1$, it can be 
straightforwardly shown, by using Eqs \ref{r+_lv} and \ref{r-_lv},
that $J \sim 1/v$. These plots show that it is possible to choose the defect 
velocity and the particle density to optimize the transport in the system in 
the direction opposite to defect motion.

%\section{Density profile and current for ${\bm q \neq 0}$}

For nonzero $q$, we do not have any closed form analytical solution of 
Eq. \ref{evolution1}. However, it can be shown from the microscopic 
dynamics that the density profile $\rho(x,t)$ 
satisfies the diffusion equation  
\be
\frac{\partial \rho}{\partial t} = D \frac{\partial^2 \rho}{\partial x^2}
\ee 
with boundary condition  for current $- D \partial \rho/\partial x = \rho_+v$ 
at $x = vt$ where the diffusion
coefficient $D=q/2$. The density profile $\rho(x,t)$ then has the solution 
\be
\rho(x,t) = \rho_{_+} e^{-(x-vt)/\xi } + \rho,
\ee 
with $\xi = D/v$. In Fig. \ref{rho_pm}(C), we plot the density as a function 
of $x$ and, in the inset, the length-scale $\xi$ as a function of $v$,
which agrees remarkably well with the above form of $\xi(v)$.
In other words, for large $v$
when the defect movement is much faster compared to the other relaxation 
time scales, $\xi$ is small and the structure 
of the density profile remains almost same as in the case of $q=0$, 
i.e., there is a bump ($\rho_+$), a trough ($\rho_-$) and almost uniform 
bulk-density. However, for small $v$, $\xi$ becomes large and 
the density profile shows extended spatial structure. Naturally, the
description of density profile in terms of only bump and trough does 
not remain valid anymore.

For large $v$, to a good approximation, $\rho_\pm$ remains 
independent of $q$ (see Fig. \ref{rho_pm}(A)). To calculate the 
current in mean-field approximation, we note that, for $q \ne 0$, the 
following three bonds contribute to the current during the time-interval 
$\tau$. 
The mean-field current across the bond between sites $\alpha-1$ and 
$\alpha$ is $\tilde q [\rho(1-\rho_-) - \rho_-(1-\rho)]$, between sites 
$\alpha$ and $\alpha+1$ is $- \tilde p \rho_+(1-\rho_-)$ and
between sites $\alpha+1$ and $\alpha+2$ is $\tilde p \rho_+(1-\rho)$. 
For large $v$, the effective rates $\tilde{p}=(v/L)(1-e^{-p/2v})$ 
and $\tilde{q}=(v/L) (1-e^{-q/2v})$ are the hopping probabilities from the 
defect and the bulk site, respectively, to the unoccupied nearest neighbor 
during the residence time $\tau$. Therefore we obtain the net current
\be
J_q(\rho, v) \simeq \tilde{p} \rho_+(\rho_--\rho)
+ \tilde{q} (\rho-\rho_-).
\label{Jq2}
\ee
Clearly, the first term is always negative, and the second term is always
positive. The competition between these two terms results in interesting 
effects like polarity reversal of current as $\rho$ and $v$ are varied.
In Fig. \ref{j_rho}(A), the current is plotted as a 
function of $\rho$. Evidently, there is no particle-hole symmetry and 
current switches sign as a function of $\rho$, with a
positive and a negative peak  in the current-density plot. 
We obtain quite good agreement between our mean-field predictions 
and simulations. Small discrepancies can be attributed to the presence 
of spatial correlations in the system.

In Fig. \ref{j_rho}(B), current is plotted as a function of $v$ for 
various values of $q$. One striking aspect in this plot is the noticeable
variation of current over almost five decades of $v$. 
For large $v \gg 1$, current decays as $1/v$, as follows
from a straightforward analysis of Eq. \ref{Jq2} where one expands $\rho_\pm$,
$\tilde p$ and $\tilde q$ in leading order of $1/v$.
For intermediate values of $v$, current shows polarity reversal, i.e.,
for any nonzero $q$ and $0<\rho<1$, there 
exists a $v_c(q,\rho)$ such that $J>0$ for $v<v_c$ and $J<0$ for 
$v>v_c$. Moreover, for any given $\rho$ and non-zero $q$ below a particular
 value, there are positive and negative 
peaks of current at particular values of $v$, indicating that most efficient 
transport can be achieved in either direction along the ring. Our
analytical results capture these broad features quite well. Particularly, 
for large $v$, the agreement between the expression in 
Eq. \ref{Jq2} and simulation is excellent. However, the quantitative 
agreement between analytical results and simulations is not good when $v$
becomes small. The closed 
form expression for $\rho_+$ at large $v$ does not remain valid anymore and 
therefore cannot be used to obtain the current in this 
regime. In the equilibrium limit of $v \to 0$, one must have $\rho_+ \to \rho$,
which is indeed the case in simulations. However, as seen from Fig.
 \ref{rho_pm}C, the decay of $\rho_+$ is extremely slow,
e.g., for $q=0.5$, over two decades of $v$, $\rho_+$ decays approximately 
by a factor of only two.

%\section{Discussion}

It should be possible to design experiments where our model 
can be realized. For example, colloidal suspension of micron-sized beads, 
naturally having excluded volume interaction, can be confined in a 
narrow channel and acted on by a moving optical tweezer, which constitutes 
a relevant experimental set up. For a typical colloidal particle of diameter 
$a = 1 \mu m$, suspended in an aqueous solution at room temperature, we 
obtain self-diffusion 
constant $D \sim 0.4 \mu m^2s^{-1}$ using Stoke-Einstein relation and the 
characteristic diffusive time-scale $4 a^2 /D \sim 2.5 s$. Our model with
 $q=0.4s^{-1}$ and packing fraction $\rho =0.5$, predicts the optimum velocity
 of the tweezer  $v \sim 2 \mu m s^{-1}$ for most efficient transport in the 
opposite direction.

In this paper, we propose a minimal but a non-trivial model to study an 
interacting-particle system driven 
by a potential barrier moving on a ring. We find that the 
particle-current has interesting nonmonotonic dependence on the velocity 
$v$ of the moving barrier and particle density $\rho$. Most remarkably, the 
current reverses its direction and even shows positive and negative peaks 
as $v$ and $\rho$ are varied separately. We have also obtained the condition 
for the optimum transport of particles, which can be achieved in both 
directions along the ring. Our analysis can be applied to the cases of a 
finite barrier ($r \ne 0$), a moving potential well ($r > p$) or multiple 
defects and could also be useful in systems
with a more complex form of interaction among the particles \cite{prep}.
From a more general perspective, our study provides important 
insights into the nature of transport in interacting-particle systems 
having a time-periodic steady state.

\appendix
\renewcommand{\theequation}{A-\arabic{equation}}
\setcounter{equation}{0}
\section{Appendix A: Calculation of $\bm {a_{\pm}}$ for $\bm {q=r=0}$ }
As defined in the main text, $a_+$ is the conditional probability that given
the defect site is occupied, the particle from the defect site hops to its
right neighbor. We use the notation $\hat{1}$ to denote an occupied defect
site and $\hat{0}$ for an empty defect site. If a particle in the defect site
has to hop to its right neighbor, then the possible local configurations are
$1\hat{1}0$ and $0\hat{1}0$. In the first case, 
the move takes place if the defect site is chosen and the particle
decides to jump to the left. The probability that this happen in the first
time-step $dt$ is $ p dt/2 $ where we discretize time in steps of infinitesimal 
interval $dt$ with $Ldt=1$ and $L \gg 1$. 
If it happens in the second infinitesimal time-step, then in the
first time-step the jump did not happen (which has a probability
$(1-p dt/2)$. Therefore the probability that the jump takes
place in the second time-step is 
\be
\left(1-\frac{p dt}{2} \right)\frac{p dt}{2}.
\ee
Similarly, the probability that the jump takes place in the third time-step is 
\be
\left(1-\frac{p dt}{2} \right)^2\frac{p dt}{2}, 
\ee
and so on. Thus, the probability that the rightward jump from the defect 
site takes place in any of the 
$\tau/dt$ time-steps ($\tau=1/v$ the residence time of the defect) is given by
\bea
\frac{p dt}{2} \left[1 + \left(1-\frac{p dt}{2} \right)+\left(1-\frac{p dt}{2}
\right)^2 + \dots + \left(1-\frac{p dt}{2} \right)^{{\tau}/{dt}} \right] 
\nonumber \\
 = \frac{p dt}{2} \frac{1-(1-p dt/2)^{\tau/dt-1}}{1-(1-p dt/2)} = 
\left(1-e^{-p/2v} \right) 
\mbox{~~~~~} 
\eea
Using similar arguments, one can show that, for the local configuration
$0\hat{1}0$, the probability that the rightward jump from the defect 
site takes place in any of the 
$\tau/dt$ time-steps is $(1-\exp(-p/v))/2$. Therefore the
expression for $a_+$ becomes
\be
a_{+} = \left[ \frac{\langle \eta^{(\alpha)}_{\alpha} 
\eta^{(\alpha)}_{\alpha+1} (1-\eta^{(\alpha)}_{\alpha+2})\rangle}{\kappa_1 
\langle \eta^{(\alpha)}_{\alpha+1}\rangle} + \frac{\langle 
(1-\eta^{(\alpha)}_{\alpha})\eta^{(\alpha)}_{\alpha+1}
(1-\eta^{(\alpha)}_{\alpha+2}) \rangle}{2 \kappa_2 \langle 
\eta^{(\alpha)}_{\alpha+1}\rangle} \label{eq:a+} \right]
\ee 
In a similar way, the expression for $a_-$ can also be derived.

The structure of the rate matrix $\mathcal W$ depends on the position of the
defect site $\alpha$ and its elements can be written in terms of $a_\pm$.
 For example, when $\alpha=1$, the matrix is
\[
{\cal W}^{(1)}= \left[ \begin{array}{cccccc}
        (1-a_+-a_-) & a_+ & 0 & \dots & 0 & a_- \\
        0 & 1 & 0 & 0 & \dots & 0 \\
        \dots & \dots & \dots & \dots & \dots & \dots \\
        \dots & \dots & \dots & \dots & \dots & \dots \\
        0 & \dots & 0 & 0 & 1 & 0 \\
        0 & 0 & \dots & 0 & 0 & 1 \\
        \end{array}
\right] \mbox{~~~~~~~~~~~~~~~~}
\]

and for $\alpha=2$

\[
{\cal W}^{(2)}=
\left[ \begin{array}{cccccc}
        1 & 0 & 0 & \dots & 0 & 0 \\
        a_{-} & (1-a_{+}-a_{-}) & a_{+} & 0 & \dots & 0 \\
        0 & 0 & 1 & 0 & \dots & 0 \\
        \dots & \dots & \dots & \dots & \dots & \dots \\
        0 & \dots & 0 & 0 & 1 & 0 \\
        0 & 0 & \dots & 0 & 0 & 1 \\
        \end{array}
\right] \mbox{.~~~~~~~~~~~~~~~~}
\]

%%%%%%%%%%%%%%%%%

%%%%%%%%%%%%
\appendix
\renewcommand{\theequation}{B-\arabic{equation}}
\setcounter{equation}{0}
\renewcommand{\thefigure}{B-\arabic{figure}}
\setcounter{figure}{0}

\section{Appendix B: Simulation Details}
\label{sec4}

The Monte Carlo simulations have been performed using the following algorithm
with random sequential update rules. We start with an initial configuration 
where $N$ randomly chosen sites of a periodic lattice of sixe $L$ are filled 
with particles. Initially we chose a particular site as the defect site which
has a different hopping rate than the rest of the system. In Fig. 
\ref{model_fig}, the oval shaped site is the defect site and $p,q,r$ are the
hopping rates for different sites. We perform the following steps 
repeatedly in the simulations.
\\
\\
\noindent {\it Step 1} - A site is
chosen at random and updated as per the transition rates shown in Fig.
\ref{model_fig}. A single Monte Carlo step (MCS) is defined as $L$ such 
update trials. 
\\
\\
\noindent {\it Step 2} - The defect site moves on the lattice with velocity 
$v$, i.e., the defect stays at a particular site for a residence time 
$\tau=1/v$ MCS. After $\tau$ MCS, the defect moves to the next site. 
\\
\\
After the system reaches the time-periodic steady state, the quantities such 
as density profile, $\rho_{\pm}$ and particle current  are measured.

\end{document}